\documentstyle[11pt,newpasp,twoside,epsf]{article} 
\def\edcomment#1{\iffalse\marginpar{\raggedright\sl#1\/}\else\relax\fi} 
\reversemarginpar 

\begin{document} 
\title{Is NGC6752 hiding a double black hole binary in its core ?}

\author{ Monica Colpi, Alessia Gualandris} 
\affil{Department of Physics, University of Milano Bicocca, Piazza della
Scienza 3, I-20126 Milano, Italy}
\author{Andrea Possenti} 
\affil{Oss. Astronomico di Bologna, Via Ranzani 1, I-40127
Bologna, Italy}

\begin{abstract} 
NGC6752 hosts in its halo  PSR J1911-5958A, a 
newly discovered binary millisecond pulsar
which is the most distant pulsar 
ever known from the core of a globular cluster.  
Interestingly, its recycling history seems in conflict  
with  a scenario of ejection 
resulting from ordinary stellar dynamical encounters.
A scattering event off a binary
system of two black holes with masses in the range of
$3-50\,M_{\odot}$ that propelled PSR~J1911-5958A
into its current peripheral orbit seems more likely.
It is still an observational challenge to unveil  
the imprint(s) left  from such a dark massive binary
on cluster's stars: PSR J1911-5958A may be the first 
case.

\end{abstract}

\section {Binary black holes in globular clusters}

It is known that a significant number of black holes (BHs) of stellar
origin  
should form
in the first million years in globular clusters (GCs),  
with masses $>3 M_{\odot}$. These BHs segregate toward the GC cores by dynamical
friction where they form double binaries [BH+BH] and multiple systems.
With time, BH$-$[BH+BH] exchange interactions become
overwhelmingly important and most of the BHs are ejected by recoil
(Sigurdsson \&
Hernquist 1993).
The last [BH+BH] binary(ies) however will have no other BHs to eject and
will remain in the cluster's core (Portegies Zwart \& McMillan 2000).
In this cosmic dance some of the interacting BHs can grow in mass 
up to $\sim 100\,M_{\odot}$ by cannibalizing other BHs (Miller \& Hamilton 2002)
and these will be preferentially retained in the GC due to their higher
inertia (a high mass BH could also be primordial, born from the heaviest
stars). 
Figure 1 shows, for a binary with masses [$m_{\rm BH}+ M_{\rm BH}\sim 50 M_{\odot}$], 
the two lower bounds on the separation $a$, denoted with $a_{\rm rec}$ and $a_{\rm
GW}$, as a function of $m_{\rm BH},$  
below which recoil by an incoming BH of 10 $M_{\odot}$ (or of $m_{\rm BH}$), and
gravitational wave (GW) emission would eject the binary and cause
coalescence, respectively.
Can such a massive dark [BH+BH] binary in a GC be detected? 
GCs host a number of millisecond pulsars (MSPs) tracing the population
of neutron stars (NSs) recycled to short periods in accretion events, after 
capturing the mass-losing  star in dynamical encounters.
We here show that PSR~J1911-5958A, a canonical
binary MSP, can probe 
the existence  of a [BH+BH] binary in NGC6752. 

\begin{figure}

\plotfiddle{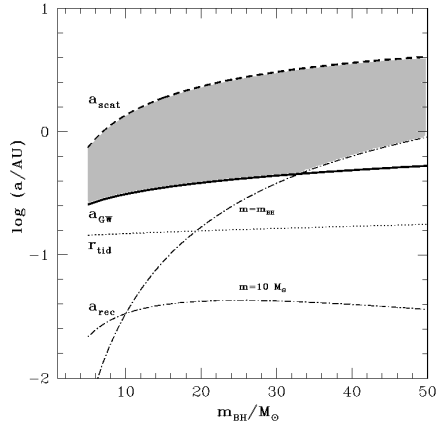}{5.truecm}{0}{30}{30}{-100}{-15}

\caption{The shaded area (between $a_{\rm scat}$ and $a_{\rm {GW}}$)
indicates the values of the [BH+BH] 
separation $a$ necessary for ejection of
PSR~J1911-5958A, and consistent with BH binary survival. $r_{\rm {tid}}$
denotes the distance below which the binary 
PSR~J1911-5958 would have been tidally disrupted by the BH tidal field
during a scattering event.}

\end{figure}

\section{A millisecond pulsar as tracer of a double black hole binary } 

PSR~J1911-5958, recently discovered in NGC6752 (D'Amico et al., 2002), is a binary
MSP found at more than 3.3 half-mass radii from the cluster core.
Its companion is a dwarf star of mass $m^*\sim 0.2M_{\odot}$ moving
along a circular orbit with  period of 0.86 days.
To reach it current position in the cluster halo, the pulsar should have been
propelled out at a speed $\approx 30$ km s$^{-1}$ close
to escape, due to recoil in an exchange
interaction [star, star]+NS $\to$ [star, NS] + star.
Using knowledge of the binary pulsar parameters and
constraints imposed by the recycling scenario, it is found that
momentum conservation during the exchange 
implies an initial orbital separation 
suspiciously close to the currently observed value, and, most notably, 
occurrence times for the exchange much longer than 
the lifetime of NGC6752 (see Colpi, Possenti,
\& Gualandris 2002, for details). 
Thus, only a massive binary can be
powerful
enough to propel PSR~J1911-5958 out from the core by transferring 
a  fraction of its binding energy to kinetic energy of the binary
pulsar, in a pure scattering event (see Figure 1). 
Nearly $3000$ stars would interact with the [BH+BH]
over a Hubble time: 
Scaling to the number of NSs and the relative fraction of active MSP, we infer an
interaction rate for  PSR~J1911-5958 with [BH+BH] of $0.5-1$ Gyr$^{-1}$ implying a
detection probability of $\sim 50\%$ considering the typical lifetime of the
pulsar.


\begin{references} 




Colpi, M., Possenti, A., \& Gualandris, A. 2002, \apj, 570, L85


D'Amico, et al. 2002, \apj, 570, L89

Miller, M.C., \& Hamilton, D.P. 2002, \mnras, 330, 232

Portegies Zwart, S.F., \& McMillan, S.L.W. 2000, \apj, 528, L17

\end{references}
\end{document}